\documentclass[pra,preprint,tightenlines,showpacs,superscriptaddress,twocolumn,10pt]{revtex4} 
\usepackage{epsfig} 

\begin{document}  
    
\title{ Excitation of highly charged hydrogen-like ions by the impact of equivelocity 
electrons and protons: a comparative study }
 
\author{ B. Najjari }   
\affiliation{ Institut Pluridisciplinaire Hubert Curien, 
Universit\'e de Strasbourg, \\ 
23 rue du Loess, BP 28, 67037 Strasbourg Cedex 2, France } 
\email{bennaceur.najjari@iphc.cnrs.fr}

\author{ A.B.Voitkiv  }   
\affiliation{ Max-Planck-Institut f\"ur Kernphysik, 
Saupfercheckweg 1, D-69117 Heidelberg, Germany } 
\email{alexander.voitkiv@mpi-hd.mpg.de}

%\date{\today} 

\begin{abstract} 

We consider excitation of highly charged 
hydrogen-like ions by the impact of equivelocity  
electrons and protons. 
The kinetic energy of the protons is 
more than three orders of magnitude larger 
than that of the equivelocity electrons. 
It is shown, however, that despite this fact, 
the electrons can be much more effective 
in inducing excitation at collision velocities 
(slightly) above the theshold for electron impact excitation. 
The basic reason for this is the strong distortion 
of the motion of the electron by the attractive field 
of the nucleus of the highly charged ion.   

\end{abstract} 


\maketitle 

%% body of paper begins here 

%\newpage 
 
\section{Introduction} 

Excitation of highly charged ions by the impact of 
charged particles (projectiles) is an interesting physical process 
which may also have many applications. In particular, 
in case of electron projectiles the studies of this process 
are of importance for 
the physics of high-temperature plasmas 
produced in laboratories and existing in astrophysical sources. 

Collisions of energetic highly charged ions with atoms represent 
an important field of research at modern accelerators of heavy ions. 
In collisions of a highly charged ion with neutral atoms the ion 
can also be excited. If the momentum transferred to the atom in 
the collision is much larger than the typical momenta of 
the electrons in the atom the excitation process can be regarded as 
occurring due to the incoherent interactions of the electron of the ion 
with the nucleus and the electrons of the atom which behave with respect 
to each other as (quasi-) free particles \cite{we}. 
In the rest frame of the ion the excitation then can be viewed as 
induced by the incoherent impacts of the ``independent'' 
beams of the atomic nucleus and atomic electrons. 

If the collision velocity is much larger than the Bohr velocity in 
the $K$-shell of the atom, the contribution $\sigma_{N}$ to the excitation cross section 
caused by the interaction with the nucleus of the atom is 
very simply related to the cross section $\sigma_{p}$ for excitation 
by proton impact: $\sigma_{N} = Z_A^2 \sigma_{p}$, where $Z_A$ is 
the charge of the atomic nucleus. 
 
We thus see that the excitation of a highly charged ion in collisions 
with atoms can, under certain conditions, be reduced to two basic processes: 
excitations in collisions with an equivelocity electron and proton.     
In this respect a question arises 
about the relative effectiveness of these two types of projectiles 
in producing the excitation. Note that althought excitation of ions 
by electronic and protonic (nuclei) projectiles has been studied 
(see e.g. \cite{fontes1}-\cite{abv-buch} and references therein), 
to our knowledge these studies were always done separately 
for electrons and protons. 

From the perspective of atomic physics the differences between 
the electrons and protons mainly include i) the huge difference in masses 
and also ii) the opposite sign of their charges. 
For instance, due to the first point, the kinetic energies of 
equivelocity electrons and protons differ 
by about a factor of $2000$. 

A trivial consequence of this fact is that in the range 
of impact velocities $v$ below the threshold velocity $v_{th}$ for incident electrons, 
where these electrons do not have enough energy to excite the ion, 
the protons do have and are capable of producing excitation.
Besides, it is also quite natural to expect that at sufficiently high impact 
velocities, where the kinetic energy of the incident electron 
is much larger than the excitation energy of the ion, the cross sections 
for excitation by equivelocity electrons and protons will converge. 

What, however, can can say about the relative effectiveness of these two projectiles 
in case when the impact velocity is already above the threshold velocity $v_{th}$ 
but the kinetic energy of the incident electron is not yet much larger than 
the excitation energy? Below in this article, where excitation 
of highly charged hydrogen-like ions by the impacts of equivelocity 
electrons and protons is considered for a broad range 
of the atomic numbers of the ions, we shall address this question.  
   
The article is organized as follows. In the next section 
we briefly outline the basic physics of proton-ion and electron-ion 
collisions and discuss how one can calculate the corresponding 
excitation cross sections. In section III we present results  
for excitation cross sections of hydrogen-like ions of nickel, 
xenon, erbium, bismuth and uranium (the corresponding atomic numbers are: 
$Z_I = 28$, $54$, $68$, $83$ and $92$). 
The main results are summarized in section IV.       

Atomic units ($\hbar = m_e = |e| =1 $) are used throughout the paper 
except where otherwise stated.   

\section{ Theory }   

To an excellent approximation the nucleus of the ion,  
which has a charge $Z_i$ ($Z_i \gg 1$), can be regarded
as infinitely heavy and the field which it creates 
as an external field.  
We shall consider the collisions in the rest frame 
of this nucleus and choose its position as the origin. 

The transition amplitude 
for the excitation of the ion by the projectile 
(an electron or a proton)
can be written according to (see, e.g. \cite{abv-buch}) 
\begin{eqnarray}
S_{fi}= - \frac{i}{c^2} \int d^4x \int d^4y \,%
j_{\mu}(x) \, D^{\mu \nu}(x-y) J_{\nu}(y). 
\label{e1}
\end{eqnarray}
Here, $j_{\mu}(x)$ and $J_{\nu}(y)$ ($\mu, \nu=0,1,2,3$) 
are the electromagnetic transition 4-current densities 
generated by the electron of the ion 
at a space-time point $x$ and by the projectile 
at a space-time point $y$, respectively,
and $D^{\mu \nu}(x-y)$ is the propagator 
of the electromagnetic field 
which transmits the  
interaction between these particles. 
The contravariant $a^{\mu}$ and 
covariant $a_{\mu}$ $4$-vectors are given by
$a^{\mu}=(a^0,{\bf a})$ and $a_{\mu}=(a^0,-{\bf a})$. 
The metric tensor $g_{\mu \nu}$ 
of the four-dimensional space is defined by 
$g_{00} = -g_{11}=-g_{22}=-g_{33}=1 $ 
and $g_{\mu \nu}=0 $ for $\mu \neq \nu$. 
In (\ref{e1}) the summation  
over the repeated greek indices is implied.       
 
Provided the transition currents in (\ref{e1}) 
(and in the corresponding exchange contribution to the amplitude)
are evaluated using 
the relativistic description of the bound and free particles, 
the treatment of the excitation process is fully relativistic. 
In particular, in this treatment there is no upper limit 
on the collision energy and also excitation 
of most heavy ions may be considered.  

\subsection{ The effect of the field of the nucleus of the ion on 
the motion of the incident and scattered particle }    

Since we suppose that the nucleus of the ion has a high charge, 
its field can in general strongly influence not only the motion of 
the bound electron but also that of the incident (and scattered) particle.  
A simple estimate for the magnitude of the effect of this field on the motion 
of the incident electron and/or proton in the process of excitation 
can be obtained in the following way \cite{we}. 
Assume that there is a particle with a charge $q$ and mass $m$ 
which is incident with a velocity $v$ on the nucleus $Z_i$. 
One can estimate the effect of the field 
by using the ratio $\varsigma = \delta p/p_i$, where  
$p_i = m \gamma v$ ($\gamma = 1/\sqrt{1-v^2/c^2}$) 
is the initial momentum of the incident particle and 
$\delta p $ is the change in the momentum of this particle 
caused by the field of $Z_i$. This change 
is roughly given by $\delta p \sim Z_i q/(bv)$, where 
$b$ is the impact parameter. For the problem of excitation 
the typical impact parameters are of the order of $1/Z_i$ or larger. 
Therefore, 
\begin{eqnarray} 
\varsigma \simeq \frac{ |q| }{ m \gamma } \frac{ Z_i^2 }{ v^2 }. 
\label{effect}
\end{eqnarray}  
In order to make the process of excitation in collisions 
with electrons energetically possible,   
one roughly needs $ Z_i^2 /m_e v^2 \stackrel{<}{\sim} 1$. 
Therefore, it follows from (\ref{effect}) 
that for the impact velocities of interest for the present article 
the parameter $\varsigma$ is very small in the case of proton projectiles 
($ \varsigma \stackrel{<}{\sim} 1 / m_p \gamma < 10^{-3} $).  
As a result, the field of the ionic nucleus does not affect 
the motion of the proton which can be regarded 
in the initial and final states as a free particle. 
In contrast, for electron projectiles $\varsigma$ may be close 
to $1$ ($ \varsigma \stackrel{<}{\sim} 1 / m_e \gamma \stackrel{<}{\sim} 1  $) 
which means that the field of the nucleus can very strongly 
distort the motion of the electron. 
Indeed, it will be seen below that this distortion 
can have a crucial impact on the process of excitation. 

\subsection{ Excitation in collisions with protons }   

The treatment of excitation of a highly charged 
hydrogen-like ion by protons is based on 
the following main points (see e.g. \cite{we}).  

First, the charge of the proton is much smaller than 
that of the highly charged nucleus of the ion. 
As a result, the interaction between the proton and the electron 
of the ion in the process of excitation is much weaker 
than the interaction between the electron and the ionic nucleus. 
Therefore, it can be regarded as a weak perturbation and may be 
taken into account within one-photon exchange 
(first-order perturbation theory). 

Second, as was already mentioned in the previous subsection, 
due to the relatively heavy mass of the proton 
the distortion of its motion caused by 
the field of the nucleus of the ion can be ignored. 
Then, regarding the proton as a Dirac particle,  
one can approximate the initial and final states of the proton 
by (Dirac) plane-waves. 

\subsection{ Excitation in collisions with electrons }

Let us now briefly discuss the treatment of 
excitation of a highly charged hydrogen-like 
ion in collisions with electrons 
(see e.g. \cite{fontes1}-\cite{moors}, \cite{we}). 

Like in the case of collisions with protons, 
the interaction between the incident electron 
and the electron of the ion is comparatively very weak. 
Therefore, the description of this interaction in the excitation process 
can be reduced to just single-photon exchange between the electrons. 

However, in contrast to the excitation by protons, 
the interaction between the incident electron and the ion 
in general cannot be treated within 
the first-order perturbation theory. The reason is that 
the motion of the incident (and scattered) electron 
can be very substantially affected by its interaction 
with the nucleus of the ion.  
This point can be addressed by describing 
not only the bound but also the continuum electron as 
moving in the Coulomb field of the nucleus of the ion. 

Further, the electrons are indistinguishable and, 
therefore, the exchange effect has to be taken into account by  
including an additional diagram (the so called exchange diagram) 
into the treatment of electron-impact excitation.   
 
Below we shall refer to the treatment, which (i) describes 
the continuum electron as moving in the Coulomb field 
of the nucleus of the ion, (ii) takes into account the interaction 
between the continuum and bound electrons within first-order perturbation 
theory and (iii) includes the exchange effect, as Approach I. 

In addition, in the next section we shall present results for 
electron impact excitation obtained by using another -- simplified -- treatment. 
This simple treatment -- termed Approach II -- 
also describes the interaction between the free and 
bound electrons within first-order perturbation theory 
but neglects the distortion of the continuum electron states 
by the field of the ion (approximating them by Dirac plane waves)  
and does not take into account the exchange effect between 
the free and bound electrons. 

Note that both these approaches do not take into account 
the channel of resonance excitation. This channel 
may become effective when the energy of the initial 
configuration of the electrons (the incident electron 
plus the electron bound in the ground state of a hydrogen-like ion) 
closely matches an energy of 
a doubly excited bound state of the correspoding helium-like ion. 
Under such conditions the excitation of a hydrogen-like ion 
may proceed via formation of a doubly excited bound state of 
the corresponding helium-like ion 
which then decays due to autoionization 
into an excited state of the hydrogen-like ion 
(see e.g. \cite{alfred_mu}).  

\section{ Numerical results and discussion }

\begin{figure}[t]
\vspace{-0.75cm} 
\begin{center}
\includegraphics[width=0.55\textwidth]{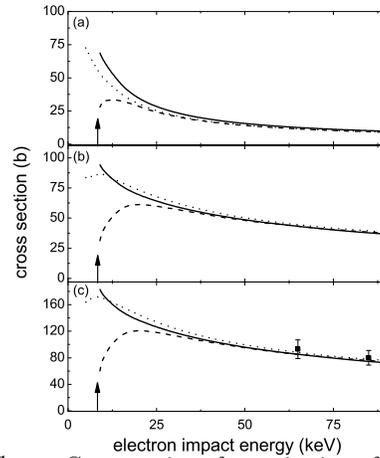}  
\end{center}
\vspace{-1.5cm} 
\caption{ \footnotesize{ Cross sections 
for excitation of hydrogen-like nickel ($Z_I=28$) by equivelocity 
electrons and protons given as a function of the electron kinetic energy. 
Sections (a), (b) and (c) show the cross sections for 
the $ 1s_{1/2} $ $ \to $ $ 2s_{1/2} $, $ 1s_{1/2} $ $ \to $ $ 2p_{1/2} $    
and $ 1s_{1/2} $ $ \to $ $ 2p_{3/2} $ transitions respectively.  
Solid and dash curves show the results for excitation by electron impact 
obtained by using Approach I and Approach II, respectively.  
Dot curve displays the results for excitation by protons. 
Experimental data for electron impact excitation 
are from \cite{thorn}. } } 
\label{nickel}   
\end{figure}

\begin{figure}[t]
\vspace{-0.75cm} 
\begin{center}
\includegraphics[width=0.55\textwidth]{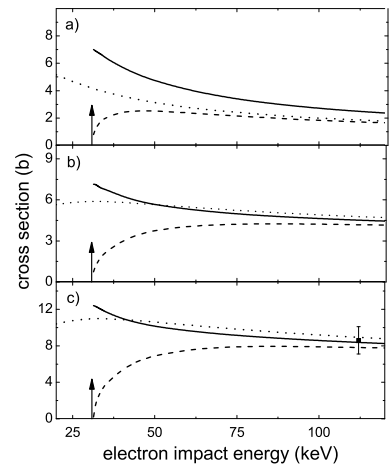}  
\end{center}
\vspace{-1.cm} 
\caption{ \footnotesize{ Same as in figure \ref{nickel} but 
for excitation of hydrogen-like xenon ($Z_I=54$). 
Experimental result for electron impact excitation 
is from \cite{xenon}. } } 
\label{xenon}   
\end{figure}

\begin{figure}[t]
\vspace{-0.75cm} 
\begin{center}
\includegraphics[width=0.55\textwidth]{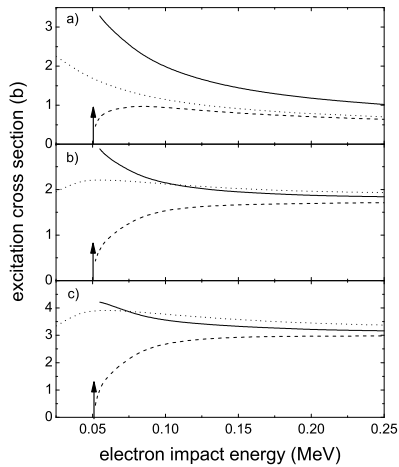}  
\end{center}
\vspace{-1.0cm} 
\caption{ \footnotesize{ Same as in figure \ref{nickel} but 
for excitation of hydrogen-like erbium ($Z_I=68$).} } 
\label{erbium}   
\end{figure}

\begin{figure}[t]
\vspace{-0.75cm} 
\begin{center}
\includegraphics[width=0.55\textwidth]{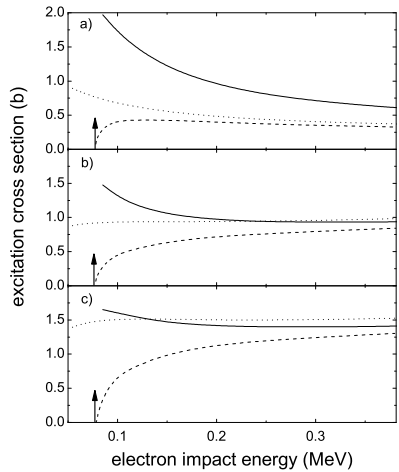}  
\end{center}
\vspace{-1.cm} 
\caption{ \footnotesize{ Same as in figure \ref{nickel} but 
for excitation of hydrogen-like bismuth ($Z_I=83$). } } 
\label{bismuth}   
\end{figure}

\begin{figure}[t]
\vspace{-0.75cm} 
\begin{center}
\includegraphics[width=0.55\textwidth]{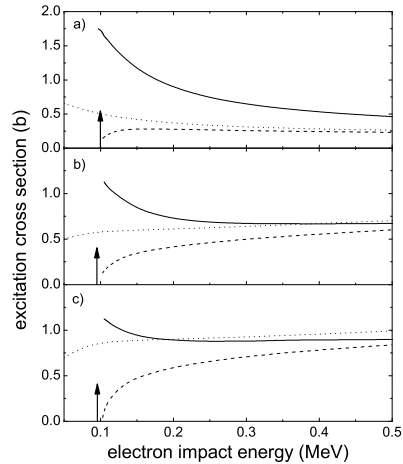}  
\end{center}
\vspace{-1.cm} 
\caption{ \footnotesize{ Same as in figure \ref{nickel} but 
for excitation of hydrogen-like uranium ($Z_I=92$).} } 
\label{uranium}   
\end{figure}

Here we shall consider excitation of 
Ni$^{27+}$(1s), Xe$^{53+}$(1s), Er$^{67+}$(1s), 
Bi$^{82+}$(1s) and U$^{91+}$(1s) ions 
caused by the impacts of equivelocity electrons and protons. 
We restrict ourselves to the excitation into the $L$-shell 
only (for which the cross sections are much larger than 
for the higher shells).  
The corresponding results are shown in figures 
\ref{nickel} - \ref{uranium} where the calculated 
cross sections for excitation by electron impact are displayed 
by solid and dash curves and those for excitation by protons 
are depicted by dot curves. 

The main conclusion, which can be drawn from the figures, 
is that, despite the huge difference in kinetic energy, 
the electrons are overally not less effective 
than protons in inducing the excitation. 
One more important point following from the figures 
is that the relative effectiveness  
of the electron projectiles compared to that of the protons  
substantially increases when the atomic number of the ion grows. 
 
The very large difference in kinetic energies 
between equivelocity electrons and protons 
makes the phase space for the final states of 
the outgoing electron 
(the cross section is proportional to the volume of this space)  
much smaller compared to that of the scattered proton. 
Since the volume of this space is proportional 
to $ \sim k_f^2 dk_f \sim k_f \varepsilon_f d \varepsilon_f/c^2 $, 
where $k_f$ and $\varepsilon_f $ are the momentum and total  
energy of the outgoing electron, it 
becomes especially small  
when the impact energy of the incident electron 
approaches the excitation threshold. 

This is why the cross section calculated using 
the simplified Approach II, in which the incident and scattered electron 
is described by plane waves, increases from zero at the excitation threshold.  
Contrary to Approach II, however, the more sophisticated 
Approach I leads to the cross sections which have their maxima 
at the electron impact energy equal to the excitation energy \cite{data}. 

Such a behaviour is the consequence of the well known singularity 
which is present for the continuum states of an electron moving 
in an attractive Coulomb field with an asymptotic momentum $k \to 0$.  
This singularity compensates for the smallness of the phase space of 
the outgoing electron. Thus, it is the distortion 
of the motion of the unbound electron by the attractive Coulomb field 
of the ion which makes the electronic projectiles 
so effective in exciting the ion.  
This distortion is especially strong for the low-velocity electrons 
which results in the fact that at the excitation threshold 
and slighly above it the electrons can be 
even much more effective than the equivelocity protons.  

\section{ Conclusion }

We have considered excitation of highly charged hydrogen-like 
ions in collisions with equivelocity electrons and protons. 
We have shown that the electronic projectiles are not less effective 
in inducing the excitation than the protons. Moreover, 
according to our results the relative effectiveness 
of electronic projectiles increases 
when the atomic number of the ion increases. 

The differences between these two types of projectiles, 
which influence the process of excitation, lie in 
the very large differences in their masses and also in 
the sign of charge. 

The large mass of protonic projectiles in general favours 
the process of excitation.  Indeed, it furnishes a large phase 
space for the scattered proton and also strongly diminishes the effect of 
the repulsion between the proton and the nucleus of the ion enabling 
the proton to come closer to the electron of the ion (compared,  
say, to an equivelocity positron) increasing their interaction.  

The small mass of electronic projectile  
has a two-fold influence on the excitation process. 
One the one hand, compared to a proton an equivelocity electron 
has much less kinetic energy which {\it per se} would make the electrons 
substantially less effective in inducing excitation close to the threshold 
compared to equivelocity protons.  
However, due to the smallness of the electron mass 
the motion of the incident and scattered electrons 
may be very strongly affected by the field of the nucleus of the ion. 
For electrons this field is attractive and pulls in the incident electron closer to the electron of the ion which increases the chances for excitation.  

Based on the results of this study one can also make a (rather obvious) 
conclusion that at the threshold a positron projectile 
would be very inefficient in inducing the excitation because of its strong repulsion by the nucleus. In particular, in collisions with positronium 
the effect of excitation at velocities slighly above $v_{th}$ would fully come from the electron while the positron would be merely a spectator. 
In this respect it is interesting to note that such a situation  
seems to take place even in collisions of a positronium with a neutral 
atom \cite{positronium} where the repulsion effect  
is much weaker than in case of collisions with a highly charged ion.    
  
\section*{Acknowledgement} 

A.B.V. acknowledges the support from the Extreme Matter Institute EMMI.

\end{document}